\documentclass[conference,a4paper]{IEEEtran}
\IEEEoverridecommandlockouts
\usepackage{cite}
\usepackage{amsmath,amssymb,amsfonts}
\usepackage{algorithmic}
\usepackage{graphicx}
\usepackage{textcomp}
\usepackage{xcolor}
\def\BibTeX{{\rm B\kern-.05em{\sc i\kern-.025em b}\kern-.08em
    T\kern-.1667em\lower.7ex\hbox{E}\kern-.125emX}}
\begin{document}

\title{LEO Doppler Matching from Power Spectrum Data with Continuity-Based Segmentation and Multi-Position Clock Offset Estimation
\thanks{This work was supported in part by Grant RS-2024-00407003 from the ``Development of Advanced Technology for Terrestrial Radionavigation System'' project, funded by the Ministry of Oceans and Fisheries, Republic of Korea;
in part by the National Research Foundation of Korea (NRF), funded by the Korean government (Ministry of Science and ICT), under Grant RS-2024-00358298; 
in part by the Future Space Navigation and Satellite Research Center through the NRF, funded by the Ministry of Science and ICT (MSIT), Republic of Korea, under Grant 2022M1A3C2074404; 
in part by the Basic Science Research Program through the NRF, funded by the Ministry of Education, Republic of Korea, under Grant RS-2025-25425504; 
and in part by the MSIT, Korea, under the Information Technology Research Center (ITRC) support program supervised by the Institute of Information \& Communications Technology Planning \& Evaluation (IITP) under Grant IITP-2024-RS-2024-00437494.
}
}

\author{\IEEEauthorblockN{Gaeun Kim}
\IEEEauthorblockA{
\textit{School of Integrated Technology} \\
\textit{Yonsei University}\\
Incheon, Korea \\
kge1950@yonsei.ac.kr}
\and
\IEEEauthorblockN{Seunghyeon Park} 
\IEEEauthorblockA{\textit{School of Integrated Technology} \\
\textit{Yonsei University}\\
Incheon, Korea \\
seunghyeon.park@yonsei.ac.kr} 
\and
\IEEEauthorblockN{Jongmin Park${}^{*}$} 
\IEEEauthorblockA{
\textit{Future Technology Center} \\
\textit{Danam Systems Inc.} \\
Anyang, Korea \\
jmpark@danam.co.kr}
{\small${}^{*}$ Corresponding author}
}

\maketitle

\begin{abstract}
Low Earth orbit (LEO) satellite Doppler measurements extracted from passive software-defined radio (SDR) spectrum data require temporal alignment with predicted satellite trajectories for reliable satellite association.
In our previous framework, Doppler slope information was used during both segment generation and clock offset estimation, while the clock offset was estimated at a single coarse receiver position.
This study proposes a satellite matching framework based on continuity-based Doppler segmentation and multi-position clock offset estimation to reduce these dependencies.
Doppler segments are first generated using only temporal and frequency continuity, thereby separating segment generation from the slope criterion used for clock offset estimation.
The clock offset is then estimated using five coarse receiver positions, consisting of one city-level nominal position and four surrounding positions, through a two-stage procedure based on slope compatibility and a combined Doppler cost.
The aligned segments are subsequently associated with candidate satellites using a matching score based on the same Doppler slope and bias-removed RMSE metrics, and the resulting associations are evaluated through receiver localization.
Experiments using six hours of passive Starlink/OneWeb monitoring data show that continuity-based segmentation alone did not improve receiver localization compared with slope-based segmentation under single-position clock offset estimation.
However, applying the proposed multi-position clock offset estimation to the same continuity-based Doppler segments reduced the localization error from 32.99~km to 1.84~km, achieving the lowest error among the evaluated methods.
The results for the evaluated dataset show improved spatial consistency of satellite associations when multi-position clock offset estimation is applied to continuity-based Doppler segments.
\end{abstract}


\section{Introduction}

Global navigation satellite systems (GNSS) serve as fundamental infrastructure for a wide range of positioning, navigation, and timing (PNT) applications.
However, GNSS reception can be limited or unreliable in indoor environments, urban canyons, and radio frequency interference conditions, motivating independent or complementary PNT solutions \cite{Kassas22:Jammer, QuezadaGaibor22:Cloud, Kapoor17:UAV, Kang25:Enhancing, Kim25:Enhancing, Park26:LNSS}.
Approaches including three-dimensional map-based positioning and learning-based signal reception classification have also been investigated for challenging urban environments \cite{Kim26:Zonotope, Jeong24:Quantum, Kim25:Set}.
In addition, signals of opportunity have attracted increasing attention, particularly signals from low Earth orbit (LEO) satellites with the growing deployment of large LEO constellations \cite{Prol22:Position, Stock25:Survey, Pinell23:Receiver}.

Because LEO satellites orbit at lower altitudes and move at higher orbital velocities than GNSS satellites, their received signals exhibit large and rapidly varying Doppler shifts \cite{Pinell23:Receiver, Jardak22:Potential}, which provide geometric information for receiver positioning when satellite orbit information and the carrier frequency are available \cite{Psiaki21:Navigation, Khalife19:Receiver}.
For a single known satellite, the changing satellite--receiver geometry can be exploited by accumulating Doppler measurements over time \cite{Neinavaie22:Acquisition}, whereas multiple LEO satellites can provide more diverse geometries \cite{Jardak24:Leveraging, AllahvirdiZadeh25:Doppler, Kozhaya23:Multi} but require association of each observed Doppler trace with its source satellite \cite{Xu25:Joint}.
Incorrect satellite association can degrade positioning performance \cite{Orabi21:Opportunistic}, making reliable association an important prerequisite for LEO Doppler positioning.

Existing studies on LEO satellite identification and Doppler estimation have primarily relied on raw or baseband in-phase and quadrature (IQ) samples or identifiable signal-level structures \cite{Kozhaya23:Multi, Farhangian20:Multi}.
Representative approaches include analyzing Starlink synchronization and waveform structures \cite{Humphreys23:Signal}, performing Doppler acquisition and tracking \cite{Neinavaie22:Acquisition}, exploiting differential Doppler measurements \cite{Neinavaie22:Differential}, and using characteristic tones or orthogonal frequency-division multiplexing (OFDM) structures \cite{Yang23:Starlink, Khalife21:Blind}.
Although effective when signal-level information is available, continuous raw signal acquisition and processing can impose substantial storage and computational requirements for long-term monitoring.

Passive software-defined radio (SDR) monitoring systems may instead retain frequency-domain power spectra rather than raw IQ samples to reduce storage and processing requirements \cite{Rajendran18:Electrosense, Pfammatter15:Software, Sorecau23:Real}.
LEO Doppler traces can still appear in such spectrum data, but explicit satellite identifiers and detailed signal structures cannot be directly exploited.
Receiver or transmitter frequency offsets and uncertainties in the recorded timestamps can also introduce discrepancies between observed and predicted Doppler trajectories \cite{Jardak23:Practical, Park26:Clock}.
Consequently, passive spectrum data require Doppler segment generation, temporal alignment with predicted trajectories, and satellite association before being used for LEO Doppler positioning.

Our previous study \cite{Park26:Clock} proposed a blind Doppler matching framework for passive SDR power scans.
The framework generated slope-consistent Doppler segments, estimated an effective clock offset by comparing observed Doppler slopes with predictions generated from two-line element (TLE) data using Simplified General Perturbations 4 (SGP4), and matched the resulting segments to candidate satellites using multiple Doppler similarity metrics.
However, the clock offset estimation procedure retains two dependencies that motivate further investigation.
First, Doppler slope information is used both during segment generation and during clock offset estimation, so the segments supplied to the offset estimator have already been selected using a slope consistency criterion.
Second, because predicted Doppler trajectories depend on the assumed receiver position, estimating the clock offset at a single coarse receiver position makes the estimate dependent on that particular position hypothesis.

In this study, we propose a satellite matching framework with two main components: \textbf{(i) continuity-based Doppler segmentation}, which separates segment generation from the slope criterion used for clock offset estimation, and \textbf{(ii) multi-position clock offset estimation}, which reduces dependence on a single assumed receiver position.
Doppler segments are generated using only temporal and frequency continuity, and candidate clock offsets are subsequently evaluated across multiple coarse receiver position hypotheses.
The final clock offset is determined using slope compatibility and a multi-position cost combining Doppler slope difference and root mean square error (RMSE) after bias removal.
Following temporal alignment, the Doppler segments are matched to candidate satellites using complementary Doppler slope and trajectory shape information, and the resulting associations are evaluated through receiver localization.

\section{Method}
\label{sec:method}

The proposed framework consists of three sequential steps: continuity-based Doppler segmentation, multi-position clock offset estimation, and satellite matching.
Receiver localization is subsequently used only to validate the resulting satellite associations.

\subsection{Continuity-Based Doppler Segmentation}
\label{subsec:continuity}

Unlike our previous study \cite{Park26:Clock}, which used Doppler slope consistency during segment generation, the proposed method generates Doppler segments using only temporal and frequency continuity.
This separates segment generation from the slope criterion subsequently used for clock offset estimation.

For each spectrum scan, spectral peaks exceeding the noise floor by more than 3~dB are detected, where the noise floor is defined as the median of all power values in the six-hour power matrix obtained by concatenating the six one-hour spectrum files.
Temporally adjacent peaks are connected into the same segment when their time interval is less than or equal to 15~s and their frequency difference is less than or equal to 50~kHz.
Segments containing fewer than five peaks or having a duration shorter than 5~s are then removed.
For each valid segment $i$, the observed Doppler slope $a_i^{\mathrm{obs}}$ is estimated by first-order linear regression of its time--frequency samples.

The position and velocity of each candidate satellite are obtained by propagating TLE data using the SGP4 orbit propagation model \cite{Yang23:Doppler, Hsu14:Assessment}.
Given a receiver position $\mathbf{r}$, the predicted Doppler shift of satellite $j$ is calculated from the satellite--receiver range rate and the carrier frequency \cite{Hsu14:Assessment, Guo23:Instantaneous} as
\begin{equation}
f_j^{\mathrm{TLE}}(t;\mathbf{r})
=
-\frac{f_c}{c}\dot{\rho}_j(t;\mathbf{r}),
\label{eq:doppler}
\end{equation}
where $f_c$ is the reference carrier frequency, $c$ is the speed of light, and $\dot{\rho}_j$ is the range rate between satellite $j$ and the receiver.

\subsection{Multi-Position Clock Offset Estimation}
\label{subsec:multi}

A clock offset between the recorded SDR timestamps and the time axis of the predicted Doppler trajectories can cause a mismatch between the observed Doppler segments and the predicted trajectories \cite{Park26:Clock}.
The estimated clock offset is treated as an effective temporal alignment parameter rather than a calibrated receiver clock error.

In the proposed method, each candidate clock offset is evaluated across multiple coarse receiver positions,
$\mathcal{P}=\{\mathbf{r}_0,\mathbf{r}_1,\ldots,\mathbf{r}_{P-1}\}$,
to reduce dependence on a single assumed receiver position.

A total of five coarse receiver positions are used for clock offset estimation.
The city-level nominal position is set to
$\mathbf{r}_0=(53.40^\circ\mathrm{N},\,2.40^\circ\mathrm{W},\,50~\mathrm{m})$,
representative of the Altrincham area.
Four surrounding positions are then generated in a cross-shaped configuration by shifting the nominal position by \(\pm\)0.15\textdegree{} in latitude and \(\pm\)0.20\textdegree{} in longitude.
The clock offset $\tau$ is evaluated over the range \textminus{}120~s to +120~s in 1~s increments, and each sample time is shifted according to
\begin{equation}
\tilde{t}_{i,k}=t_{i,k}+\tau.
\label{eq:time_shift}
\end{equation}

The proposed clock offset estimation consists of two stages.
Stage~1 selects candidate offsets using the slope compatibility count and consensus across the coarse receiver positions, and Stage~2 refines the selected offsets using a combined cost.

\subsubsection{Stage 1: Slope Compatibility Count}
\label{subsubsec:stage1}

For each candidate receiver position $\mathbf{r}_p$ and clock offset $\tau$, the predicted Doppler samples of candidate satellite $j$ are evaluated at the shifted segment times $\tilde{t}_{i,k}$.
The corresponding predicted Doppler slope $a_{i,j}^{\mathrm{TLE}}(\tau;\mathbf{r}_p)$ is estimated by fitting a first-order linear model to $f_j^{\mathrm{TLE}}(\tilde{t}_{i,k};\mathbf{r}_p)$ over the samples of segment $i$.
The Doppler slope difference between observed segment $i$ and candidate satellite $j$ is then defined as
\begin{equation}
\Delta a_{i,j}^{(p)}(\tau)
=
\left|
a_i^{\mathrm{obs}}
-
a_{i,j}^{\mathrm{TLE}}(\tau;\mathbf{r}_p)
\right|.
\label{eq:slope_difference}
\end{equation}

Let $\mathcal{C}_i(\tau)$ denote the set of candidate satellites that temporally overlap with segment $i$ after applying candidate clock offset $\tau$.
Specifically, a satellite is included in $\mathcal{C}_i(\tau)$ when the segment interval after applying the candidate clock offset overlaps with its visibility interval, expanded by 90~s at both ends, based on the available transit information.
Segment $i$ is considered slope compatible if at least one candidate satellite satisfies $\Delta a_{i,j}^{(p)}(\tau)<\gamma_a$, where $\gamma_a=150~\mathrm{Hz/s}$ is used in this study.
The number of slope-compatible segments at receiver position $\mathbf{r}_p$ and clock offset $\tau$ is calculated as
\begin{equation}
N_p(\tau)
=
\sum_{i=1}^{M}
\mathbf{1}
\left[
\min_{j\in\mathcal{C}_i(\tau)}
\Delta a_{i,j}^{(p)}(\tau)
<
\gamma_a
\right],
\label{eq:compatible_count}
\end{equation}
where $M$ is the total number of segments and $\mathbf{1}[\cdot]$ denotes the indicator function.

At each receiver position, offsets whose compatibility counts are equal to the maximum or lower than the maximum by one segment are retained in the locally preferred candidate set $\mathcal{T}_p$.
The one-segment margin is introduced to prevent potentially valid offsets from being prematurely excluded due to small variations in the discrete compatibility count among neighboring offsets.
The position vote $V(\tau)$ is then defined as the number of coarse receiver positions that support clock offset $\tau$:
\begin{equation}
V(\tau)
=
\sum_{p=0}^{P-1}
\mathbf{1}
\left[
\tau\in\mathcal{T}_p
\right].
\label{eq:position_vote}
\end{equation}

If only one offset has the maximum position vote, it is retained as the sole candidate for Stage~2.
If multiple offsets share the maximum vote, the median of the compatibility counts $N_p(\tau)$ across receiver positions is compared among the tied offsets.
The median is used to reduce the influence of unusually high or low counts at an individual receiver position and to provide a robust measure of compatibility across the coarse receiver positions.
Offsets whose median count is equal to the maximum median count or lower by the same one-segment margin are retained in the Stage~2 shortlist $\mathcal{T}_S$.

\subsubsection{Stage 2: Combined Cost Refinement}
\label{subsubsec:stage2}

Because the compatibility count used in Stage~1 is discrete, identical or similar counts can occur for several neighboring clock offsets.
Stage~2 therefore distinguishes the shortlisted offsets by jointly considering Doppler slope difference and trajectory shape similarity.

Observed LEO Doppler traces may contain approximately constant frequency biases arising from receiver or transmitter frequency offsets \cite{Jardak23:Practical, Kozhaya25:Unveiling}.
Let $f_{i,k}^{\mathrm{obs}}$ denote the observed Doppler frequency of the $k$th sample in segment $i$, referenced to the carrier frequency $f_c$.
For each candidate satellite, the mean residual within the segment is removed to reduce the effect of a constant frequency bias.
The residual between segment $i$ and candidate satellite $j$ is defined as
\begin{equation}
e_{i,j,k}^{(p)}(\tau)
=
f_{i,k}^{\mathrm{obs}}
-
f_j^{\mathrm{TLE}}
\left(
\tilde{t}_{i,k};
\mathbf{r}_p
\right).
\label{eq:frequency_residual}
\end{equation}

The RMSE after bias removal is then calculated as
\begin{equation}
R_{i,j}^{(p)}(\tau)
=
\sqrt{
\frac{1}{N_i}
\sum_{k=1}^{N_i}
\left[
e_{i,j,k}^{(p)}(\tau)
-
\bar{e}_{i,j}^{(p)}(\tau)
\right]^2
},
\label{eq:bias_removed_rmse}
\end{equation}
where $N_i$ is the number of samples in segment $i$ and $\bar{e}_{i,j}^{(p)}(\tau)$ is the mean residual for the corresponding candidate.
Thus, $R_{i,j}^{(p)}(\tau)$ quantifies the difference in trajectory shape between the observed and predicted Doppler traces after removing a constant frequency bias.

Because the Doppler slope difference and RMSE have different units, fixed normalization scales are used when combining the two metrics.
The combined cost for segment $i$, candidate satellite $j$, and receiver position $\mathbf{r}_p$ is defined as
\begin{equation}
J_{i,j}^{(p)}(\tau)
=
w_a
\min
\left(
\frac{\Delta a_{i,j}^{(p)}(\tau)}{A_0},
5
\right)
+
w_r
\min
\left(
\frac{R_{i,j}^{(p)}(\tau)}{R_0},
5
\right).
\label{eq:combined_cost}
\end{equation}

The normalization scales $A_0=500~\mathrm{Hz/s}$ and $R_0=2000~\mathrm{Hz}$ are adopted from our previous study \cite{Park26:Clock}, while $w_a=0.6$ and $w_r=0.4$ are fixed in this study to give slightly greater emphasis to Doppler slope consistency.
Each normalized term is clipped at 5 to prevent extreme mismatches from dominating the combined cost.

For each segment, the minimum combined cost over the candidate satellites is first obtained.
At each receiver position, the median of these segment costs is calculated, followed by another median across the coarse receiver positions.
The resulting multi-position combined cost is defined as
\begin{equation}
J_{\mathrm{MP}}(\tau)
=
\operatorname{median}_{p}
\left\{
\operatorname{median}_{i}
\left[
\min_{j\in\mathcal{C}_i(\tau)}
J_{i,j}^{(p)}(\tau)
\right]
\right\}.
\label{eq:multi_position_cost}
\end{equation}

The final clock offset is selected from the Stage~1 shortlist as
\begin{equation}
\tau^\ast
=
\arg\min_{\tau\in\mathcal{T}_S}
J_{\mathrm{MP}}(\tau).
\label{eq:optimal_clock_offset}
\end{equation}

\subsection{Satellite Matching}
\label{subsec:matching}

After applying the estimated clock offset $\tau^\ast$ to all observed segments, satellite matching is performed using the city-level nominal position $\mathbf{r}_0$ defined above.
The reference receiver coordinates are not used in either clock offset estimation or satellite matching and are used only to evaluate the subsequent receiver localization results.

For each segment $i$ and candidate satellite $j$, the Doppler slope difference $\Delta a_{i,j}$ and RMSE $R_{i,j}$ denote the corresponding metrics defined in the previous subsection, evaluated using the final clock offset $\tau^\ast$ and the city-level nominal position $\mathbf{r}_0$.
These two metrics are then converted into similarity scores using the same normalization scales as in the clock offset estimation:
\begin{equation}
q_a
=
\max\left(
0,1-\frac{\Delta a_{i,j}}{A_0}
\right),
q_r
=
\max\left(
0,1-\frac{R_{i,j}}{R_0}
\right).
\label{eq:similarity_scores}
\end{equation}

To prevent a large mismatch in one metric from being excessively compensated by high similarity in the other, an exponential penalty is applied.
Specifically, $P(x;T)=1$ for $x\leq T$ and $P(x;T)=\exp[-2(x-T)/T]$ otherwise.
To provide a sufficiently rapid decay for mismatches beyond the threshold while leaving candidates within the threshold unpenalized, the exponential decay coefficient is fixed at 2.
The combined matching score is then defined as
\begin{equation}
S_{i,j}
=
\left(
w_a q_a + w_r q_r
\right)
\min
\left[
P(\Delta a_{i,j};A_0),
P(R_{i,j};R_0)
\right].
\label{eq:matching_score}
\end{equation}

Each segment is associated with the candidate satellite having the highest $S_{i,j}$, and the resulting associations are used in the subsequent receiver localization procedure.

\section{EXPERIMENT}
\label{sec:experiment}

\subsection{Dataset}
\label{subsec:dataset}

This study evaluates the proposed method using a publicly available Starlink/OneWeb passive satellite monitoring dataset collected in the Altrincham area of the United Kingdom \cite{Watson24:Starlink, Watson23:Raw}. 
To separate the evaluation data from the time interval used in the previous study \cite{Park26:Clock}, six hours of observations collected from 03:00 to 09:00 on March 11, 2023, are analyzed. 
The dataset consists of six one-hour Kerberos spectrum files, together with the corresponding satellite transit information and Starlink and OneWeb TLE data.
The monitoring receiver remained stationary throughout the analyzed six-hour period.

The input data consist of frequency-domain power scans rather than raw IQ samples; no signal demodulation is performed, and no satellite identifier is used.
Spectral peaks and the Doppler segments generated from them are used for satellite matching, with a reference carrier frequency of 11.325~GHz for the predicted Doppler calculation \cite{Jardak23:Practical}.

\subsection{Comparison Methods}

Three methods are compared to evaluate the two components of the proposed framework.
All methods use the same satellite matching score and receiver localization procedure.

\begin{itemize}
    \setlength{\itemsep}{0pt}
    \setlength{\parskip}{0pt}
    \setlength{\parsep}{0pt}

    \item \textbf{M1}: Slope-based Doppler segmentation using the slope consistency criterion adopted from our previous framework \cite{Park26:Clock}, followed by single-position clock offset estimation at the city-level nominal position $\mathbf{r}_0$ defined in Section~\ref{subsec:multi}.

    \item \textbf{M2}: Continuity-based Doppler segmentation with the same single-position clock offset estimation at $\mathbf{r}_0$ as M1.

    \item \textbf{M3 (Proposed)}: The same continuity-based Doppler segmentation as M2, with clock offset estimation extended from the single position $\mathbf{r}_0$ to the five coarse receiver positions defined in Section~\ref{subsec:multi}.
\end{itemize}

Accordingly, the M1--M2 comparison evaluates the effect of separating Doppler segmentation from the slope criterion, whereas the M2--M3 comparison evaluates the effect of multi-position clock offset estimation.

\subsection{Receiver Localization for Validation}
\label{subsec:receiver}

The satellite associations obtained by each method are evaluated through receiver localization because LEO Doppler observations contain receiver position information through the satellite--receiver geometry \cite{Hsu14:Assessment, Guo23:Instantaneous} and can be used for nonlinear receiver state estimation \cite{Xu24:Doppler}.
Here, localization is used only to assess the spatial consistency of the selected satellite associations rather than to demonstrate a high-precision positioning algorithm.

For each method, the selected satellite for each segment is fixed before localization.
The predicted Doppler trajectory is then recomputed for each candidate grid position using the satellite states propagated from the TLE data with SGP4.
The bias-removed RMSE in Eq.~\eqref{eq:bias_removed_rmse} is evaluated by replacing the coarse receiver position $\mathbf{r}_p$ with the candidate grid position $\mathbf{r}$ and is denoted by $R_{i,j}(\tau;\mathbf{r})$; its mean over all matched segments is used as the localization cost.

The receiver position estimate for method $m$ is expressed as
\begin{equation}
\hat{\mathbf{r}}_m
=
\arg\min_{\mathbf{r}\in\mathcal{G}}
\frac{1}{M_m}
\sum_{i=1}^{M_m}
R_{i,j_i^{\ast(m)}}
\left(
\tau_m;\mathbf{r}
\right),
\label{eq:receiver_localization}
\end{equation}
where $\mathcal{G}$ denotes the localization grid, $j_i^{\ast(m)}$ denotes the satellite assigned to segment $i$ by method $m$, and $M_m$ is the number of matched segments used for localization.
The method-specific clock offset $\tau_m$ is obtained using single-position clock offset estimation for M1 and M2, and the proposed multi-position clock offset estimation for M3.

To reduce the computational load while covering a sufficiently wide search region, receiver localization is performed using a coarse-to-fine two-stage grid search. 
First, a coarse search is performed over the latitude range 49.5\textdegree--61.0\textdegree{} and longitude range \textminus{}8.5\textdegree--2.0\textdegree{} with a grid spacing of 0.5\textdegree{}.
A fine-search region extending \(\pm\)0.5\textdegree{} in both latitude and longitude is then defined around the coarse optimum obtained for each method, and the fine search is performed with a grid spacing of 0.05\textdegree{}.

The final receiver position is determined as the fine-grid point having the minimum mean bias-removed RMSE. 
The reference receiver coordinates are not included in the grid-search objective.
The localization error is defined as the horizontal distance between the estimated and reference receiver positions and is calculated only after the search is completed.

\section{RESULTS}
\label{sec:results}

\subsection{Satellite Matching Results}
\label{subsec:matching_result}

Applying continuity-based Doppler segmentation to the six-hour passive SDR monitoring dataset produced 137 valid Doppler segments, whereas slope-based segmentation produced 138 segments.

\begin{figure}[t]
    \centering
    \includegraphics[width=\columnwidth]{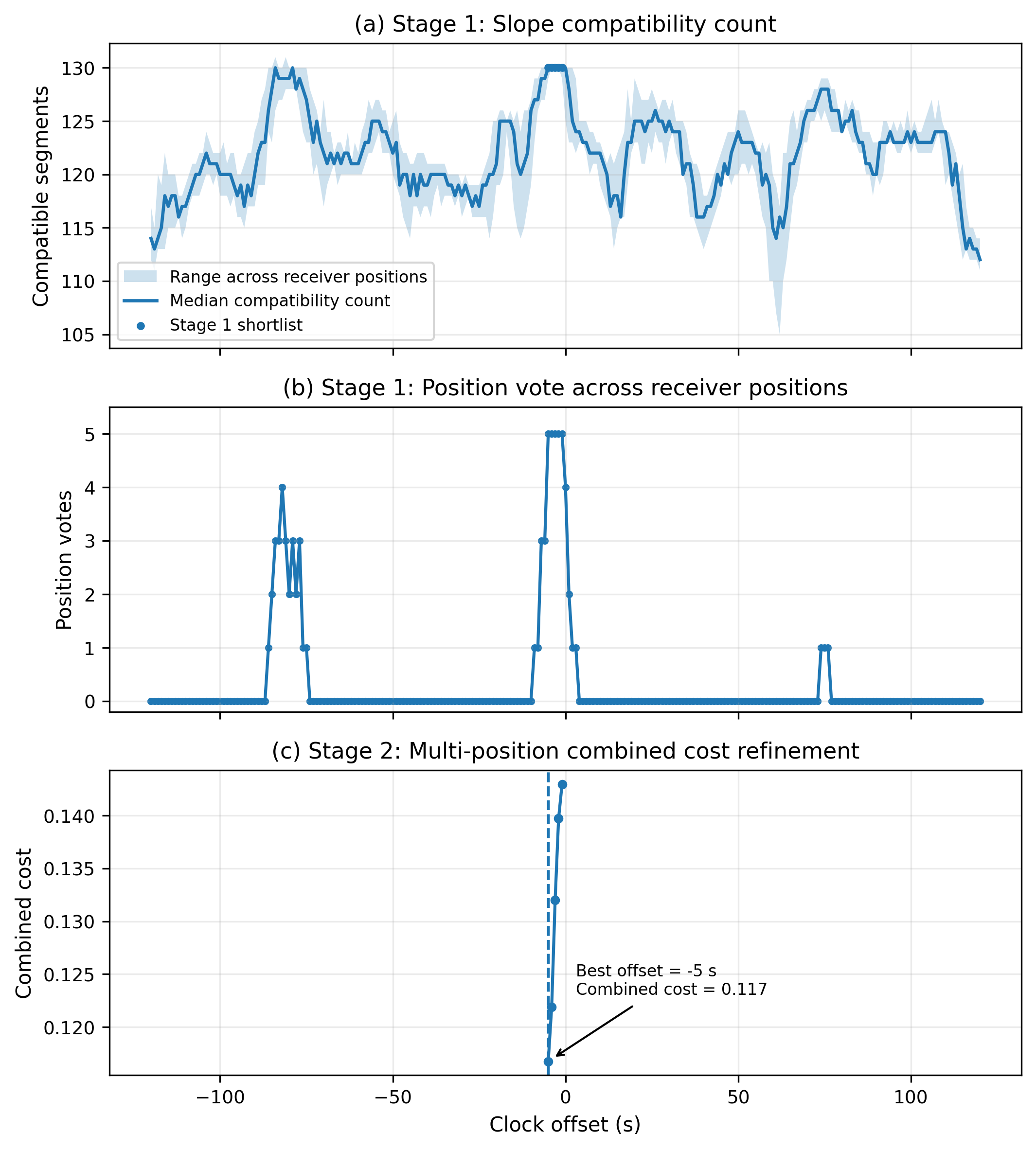}
    \caption{Results of the proposed multi-position clock offset estimation:
(a) Stage~1 slope compatibility counts across receiver positions and shortlisted offsets;
(b) Stage~1 position votes; and
(c) Stage~2 combined cost over the shortlisted offsets.}
    \label{fig:offset_search}
\end{figure}

For M1 and M2 using single-position clock offset estimation, the estimated clock offset was +22~s in both cases, with 110/138 and 105/137 slope-compatible segments, respectively.
For M3, Stage~1 achieved a maximum position vote of 5/5 and a median compatibility count of 130/137, retaining five candidate offsets from \textminus{}5 to \textminus{}1~s.
Stage~2 refinement using the multi-position combined cost selected \textminus{}5~s as the final clock offset, as shown in Fig.~\ref{fig:offset_search}.

After applying the estimated clock offset for each method, each valid Doppler segment was associated with the candidate satellite having the highest matching score.
The median Doppler slope differences for M1, M2, and M3 were 35.85, 37.80, and 29.80~Hz/s, respectively, while the corresponding median bias-removed RMSE values were 427.25, 442.50, and 402.70~Hz.
Both metrics were lowest for M3.

Because the ground-truth satellite identity of each Doppler segment is unavailable, the association accuracy cannot be determined from the matching statistics alone.
Therefore, the selected satellite associations were further evaluated through receiver localization.

\subsection{Receiver Localization Results}
\label{subsec:localization_result}

The segment--satellite associations obtained by each method were fixed and evaluated using the coarse-to-fine receiver localization described in Section~\ref{subsec:receiver}.
The final localization results are summarized in Table~\ref{tab:localization}.

\begin{table}[t]
\centering
\caption{Method configurations and receiver localization results.}
\label{tab:localization}
\setlength{\tabcolsep}{1.8pt}
\renewcommand{\arraystretch}{1.35}
\footnotesize

\begin{tabular}{c c c @{\hspace{3.6pt}} c c}
\hline
\multicolumn{3}{c}{Method configuration} &
\multicolumn{2}{c}{Results} \\
\cline{1-3}
\cline{4-5}

Method &
Segmentation &
Offset estimation &
Estimated $\tau$ &
Localization error \\
\hline

M1 & Slope-based      & Single-position & +22~s           & 21.38~km \\
M2 & Continuity-based & Single-position & +22~s           & 32.99~km \\
M3 & Continuity-based & Multi-position  & \textminus{}5~s & \textbf{1.84~km} \\
\hline
\end{tabular}
\end{table}

As shown in Table~\ref{tab:localization}, M3 achieved the lowest localization error of 1.84~km, compared with 21.38 and 32.99~km for M1 and M2, respectively.

From M1 to M2, continuity-based segmentation was introduced to avoid the repeated use of Doppler slope information in both segmentation and clock offset estimation and to reduce the dependence between the two stages.
This change alone did not reduce the localization error.
However, applying multi-position clock offset estimation to the same continuity-based Doppler segments reduced the localization error from 32.99~km in M2 to 1.84~km in M3.
For the evaluated dataset, using multiple receiver position hypotheses for clock offset estimation improved the spatial consistency of the resulting satellite associations, as reflected by the substantially lower receiver localization error.

\section{Conclusion}
\label{sec:conclusion}

This study proposed an LEO Doppler matching framework based on continuity-based Doppler segmentation and multi-position clock offset estimation for power spectrum data.
The continuity-based segmentation generates Doppler segments using only temporal and frequency continuity, thereby separating segment generation from the slope criterion used for clock offset estimation.
The clock offset is subsequently estimated across multiple coarse receiver positions using slope compatibility and a combined Doppler cost.
Experiments using six hours of Starlink/OneWeb monitoring data showed that continuity-based segmentation alone did not improve receiver localization under single-position clock offset estimation.
However, applying multi-position clock offset estimation to the same continuity-based segments reduced the localization error from 32.99~km to 1.84~km and yielded the lowest median values for both Doppler slope difference and bias-removed RMSE among the evaluated methods.
These results show that, for the evaluated dataset, incorporating multiple receiver position hypotheses into clock offset estimation improved the spatial consistency of satellite associations while allowing Doppler segment generation to remain independent of the slope criterion.
Because the present evaluation is limited to a single six-hour recording session at one location, further validation using additional passive SDR datasets collected at different locations and times is needed to assess the generality of the observed improvement.

\section*{Acknowledgment}

Generative AI (ChatGPT, OpenAI) was used solely to assist with grammar and language improvements during the manuscript preparation process.  
No content, ideas, data, or citations were generated by AI.  
All technical content, methodology, analysis, and conclusions were written and verified solely by the authors.

\bibliographystyle{IEEEtran}
\bibliography{mybibfile, IUS_publications}

\end{document}